\documentstyle[aps,eqsecnum,epsfig]{revtex}
\begin{document}
\title{Single Chain Force Spectroscopy: Sequence Dependence}
\author{Namkyung Lee \and Thomas A. Vilgis}
\address{Max-Planck-Institut f\"ur Polymerforschung,
Ackermannweg 10,
55128 Mainz, Germany}
\date{\today}
\maketitle
\begin{abstract}
  We study the elastic properties of a single A/B copolymer chain with a
  specific sequence.  We predict a rich structure in the force extension
  relations which can be addressed to the sequence.  The variational method is
  introduced to probe local minima on the path of stretching and releasing.
  At given force, we find multiple configurations which are separated by
  energy barriers. A collapsed globular configuration consists of several
  domains which unravel cooperatively.  Upon stretching, unfolding path shows
  stepwise pattern corresponding  to the unfolding of each domain.  While
  releasing, several cores can be created simultaneously in the middle of the
  chain resulting in a different path of collapse.
\end{abstract}
\section{Introduction}
Recent advances in nano manipulation allow direct access to single molecule
elasticity using Atomic Force Microscopy
(AFM) \cite{rief:97.1,kellermayer97,senden:98.1,li:98.1,courvoisier:98.1,liphardt01}.
AFM seems to be an excellent technique to study interactions in single chains
by use of their elastic response.  As a result of interplay between entropy,
interactions between monomers and environments (such as solvents, other
chains, external fields) the resulting force-extension relations allow deep
insight into the nature of the interactions and conformations. In some polymer
systems (including biopolymers and proteins), intrachain self assembly
produces  secondary or tertiary structures and the elastic responses
reflect this structural hierarchy \cite{borisov99,block,klimov99}.
Furthermore, mechanically induced folding/unfolding leads a single molecule to
undergo  structural transitions following well defined specific paths without
being frustrated. Especially in the case of proteins there is considerable
hope to gain insight into folding mechanisms and design.
 
In the present letter we study the conformation of random A/B copolymers and
their responses to external forces. 
As the response reveals information of the sequence and structure 
of the chain, this allows for ``single chain force spectroscopy''.
We use variational methods to provide some understanding of the elasticity of
polymer with specific sequence. The present model corresponds also to the most
simple ``protein model'' \cite{ramanathan94} when a polymer chain consists of
more than one constituents (here A and B monomers) which are arranged in a
specific sequence along the backbone. The sequence carries relevant
information since according to the arrangement of the monomers certain
conformations are more preferred than others.

Although the difficult nature of the problem suggests a detailed mathematical
analysis (which is partly found below), some physical scaling ideas help to
understand the problem.  Blob pictures and estimates of relevant length scales
provide an instructive physical picture for homopolymer deformation in poor
solvent condition.  Here relevant length scale is the thermal blob size
$\xi_{\rm T}$. Each blob carries energy of $k_{\rm B}T$.  A sharp structural
transition occurs when the 
Pincus blob size $\xi_P$($\equiv k_BT/f$) reaches  
to this thermal blob size $\xi_P = \xi_T$.

Experimentally, two scenarios are available: either the displacement or the
force are
controlled \cite{rief:97.1,kellermayer97,senden:98.1,li:98.1,courvoisier:98.1,liphardt01}.
When the displacement is imposed and the force is measured, the unfolding
of sequence of domains leads to repeating ``saw tooth'' pattern in force
extension curve \cite{rief:97.1,kellermayer97}.
In constant force measurement,
folding-unfolding transition appears at characteristic force, which is shown
as ``plateau'' \cite{liphardt01}. In the present study we minimize the free
energy at given force (constant force measurement).  As we will show later, we
find multiple minima at a given force.  These metastable states are created by
the heterogeneity of the sequence.  Some of these configurations are visited
depending on the history of folding/unfolding.

\section{Anisotropic Variational Method}

Formally we consider a polymer chain consisting of N monomers of size $b$ 
under  tension which is described by 
\begin{equation}
H = H_{\rm c} + H_{\rm i}  + f\cdot({\vec{r}}_N-{\vec{r}}_1).
\end{equation}
The first term corresponds to the elastic and connectivity properties of the
polymer chain and is approximated by the Wiener measure \cite{de}. 
The second term contains all interactions between monomers,
especially,  two and three body interactions of the virial expansion 
are included here. The three body
term is essential since depending on the solvent quality certain monomers may
attract each other, which results in a (partial) collapse of the chain.
The two body interactions can be attractive or repulsive depending
on the type of the monomer pairs.  The density of the collapsed globule is
determined by balance between the three body repulsions and two body
attractions.  The last term is the elastic energy term which
represents  the external
force field. The variational free energy is defined as usual by $F_V \equiv
\langle H - H_t \rangle_{0} +F_0 $ where $\langle \cdots \rangle_0$ stands for
the average over the variational probability distribution:
\begin{equation}
P_V({\vec{r}_1},..,\vec{r}_N) = Z_V^{-1} \exp \{
-H_t(\vec{r}_1,..,{\vec{r}_N}) \},
\end{equation}
where  $Z_V$ is the normalization constant  satisfying 
$\int P_V =1 $ and  $F_0=  -k_{\rm B}T \log Z_V$.   

The presence of an external force breaks the spherical symmetry.  We need,
therefore, to distinguish the deformation in parallel and perpendicular
direction with respect to the external force.  In terms of two components in
correlation function, we choose consequently the trial Hamiltonian $H_t$ as
follows
\begin{equation}
H_t({\vec r}_1, \dots ,{\vec r}_N)/k_{\rm B}T =
\frac{1}{2}\sum\limits_{j=1}^N\sum\limits_{l=1}^N
[G^{-1}_{\parallel}(j,l)({\vec r_{\parallel}})^j\cdot({\vec r_{\parallel}})^l +
G^{-1}_{\perp}(j,l)({\vec r_{\perp}})^j\cdot ({\vec r_{\perp}})^l],
\label{Ham}
\end{equation}
where $G_{\parallel}(j,l),G_{\perp}(j,l)$ are correlation functions between
$j-l$ monomers in parallel and perpendicular direction with respect to the
external force.  By minimizing the trial free energy function with respect to
$G_{\perp},G_{\parallel},$ we obtain two sets of self-consistent equations.
These self-consistent equations for both components are coupled, in contrast to
systems with spherical symmetry as studied previously \cite{jonsson93,variational1,variational2}.

The  square of the distance $b(i,j)$   between two monomers $i$ and $j$  
is related to $G_{\parallel}(i,j)$  by 
\begin{equation}
b_{\parallel}(i,j) = G_{\parallel}(i,i)+G_{\parallel}(j,j)-G_{\parallel}(i,j)-G_{\parallel}(j,i).
\end{equation}  
Therefore, the square of the mean end to end distance $\langle
R^2(1,N)\rangle$ is determined by the quantity $b_{\parallel}(1,N)$.  The
detailed derivation and solution of the self-consistent equation is shown
elsewhere \cite{afm01b}. 
This procedure corresponds  to the self-consistent
one-loop approximation \cite{feynman}.

\section{Stretching a homopolymer in a poor solvent} 
As a test run we apply the variational calculation to calculate the elastic
response of a homopolymer in a poor solvent.  This problem has been studied
theoretically (scaling \cite{halperin91}, MC simulation \cite{lai98}),
and experimentally \cite{exp}.

Using the present variational method, we confirmed the scaling picture of
force-extension relations of homopolymer.  The position of the plateau in each
curve captures the characteristic force $f_{\rm c}$ at each solvent condition
and is in agreement with scaling picture.  When we follow the local minima, we
observe a hysteresis effect at the boundary of the first order phase
transition.  Around $f_{\rm c}$, we access to local minima with different
extensions.  This means that chain configuration becomes metastable. At the
transition point, tadpole and ellipsoidal conformations coexist.

\section{Stretching a heteropolymer in a poor solvent} 
Now we consider a random A/B polymer chain of length $N$. The interaction
Hamiltonian for the pairwise interaction is given by
\begin{equation}
H_{\rm i} = 
\frac{1}{2}\sum_{i=1}^N \sum_{j=1}^N  v_{ij} \delta(\vec{r}_i -
\vec{r}_j ),
\end{equation}
where the two body interaction possesses a random nature, i.e., the
interactions are chosen as a random $N\times N$ matrix \cite{obukhov86,sfatos93,garel88}:
\begin{equation}
v_{ij} = v_0 + [\alpha(\sigma_i + \sigma_j) + \chi (\sigma_i
\sigma_j)].
\label{v2}
\end{equation}
The sequence of monomers is described by variables {$\sigma_i$}.  $\sigma_i
=-1$ if monomer $i$ is of type A (say, hydrophobic) and $\sigma_i =1$ if it is
of type B (hydrophilic).  To be more specific, a Flory like mixing parameter
$\chi = (v_{\rm AA} +v_{\rm BB})/2 -v_{\rm AB}$ is negative when similar
monomers attract each other.  The second virial coefficient $v_0 < 0$ provides
that average attraction between monomers.  Obviously, the results will depend
on the specific sequence of type A and type B monomers. For technical reasons
we assume that the average solvent quality is poor compared to the fluctuation
due to randomness ($|v_0| < \chi$), therefore the thermodynamic ground state
is always a globular state. Before a further discussion on 
the force extension properties
for given sequences, we consider the physical effects of the
randomness in our model to see what  we can  expect.

\section{Statistics}

To get a physical understanding for the problem we average over the disorder
and assume a Gaussian distribution for the disorder part in the two body
interaction, i.e., $P(\sigma_i) = e^{-\sigma_i^2/2\delta} $.  We obtain the
first order correction in the effective binary second virial coefficient
between arbitrary monomer pairs $l-m$: $\{v^{\rm eff}\}_{lm} = v_0 -
\alpha^2/\chi $ for $l\neq m$.

If the interactions between the similar monomers are  equal ($v_{\rm AA}=v_{\rm
  BB}$) then we have $\alpha = 0$. We expect no effective change in the
effective solvent quality.  In our consideration, we assume that $\alpha \sim
\chi$.  The Flory mixing parameter $\chi$ represents also the strength of
disorder.  Each thermal blob size within the collapse globular phase will be
represented by the effective second virial coefficient $v_{\rm eff}$ instead
of $v_0$, which leads to $\xi^{\rm random}_{T} = b^4/v^{\rm eff}$.  The
average thermal blob size is $\xi^{\rm random}_{T} = \xi^{\rm homo} ( 1-
\alpha/|v_0|)$.  The mean square average of $v^{\rm eff}$ is $\sim
\alpha/\sqrt{N}$.  For small disorder, for each sequence, the structural
transition occurs at large scale.  The variance in characteristic force $f_{\rm
  c}$ reflects the size of disorder ($\sim \alpha \sqrt{N}$).  Although the
mean net charge is zero, a typical random copolymer of size $N$ has excess
charge of order $\pm \sqrt{N}$.  For some sequences, there might be an excess
charge $\sum \sigma_i \neq 0$.  The characteristic force $f_{\rm c}$ is
changing accordingly due to the shift in an effective solvent quality.

Depending on the mean square fluctuation, we divide the force-extension curve
into three distinctive regime.  For small extension ($\xi_P >
\xi_T^{\rm largest}$, $f< k_{\rm B}T/\xi^{\rm largest}_T $) a collapsed
globule deforms to ellipsoidal conformation.  For large extension ($\xi_P <
\xi_T^{\rm smallest}$, $f> k_{\rm B}T/\xi^{\rm smallest}_T $), all thermal
blobs are linearly aligned.  No conformational degree of freedom is left for
rearrangement in thermal blobs.  The blob size is then simply equal to Pincus
blob size.  In intermediate regime ($\xi^{\rm smallest}<\xi_P< \xi^{\rm
 largest}$), we expect fluctuations in $z(f)$ to  depend on the
system size and external force.

\begin{figure}
\centering \includegraphics[width=12cm]{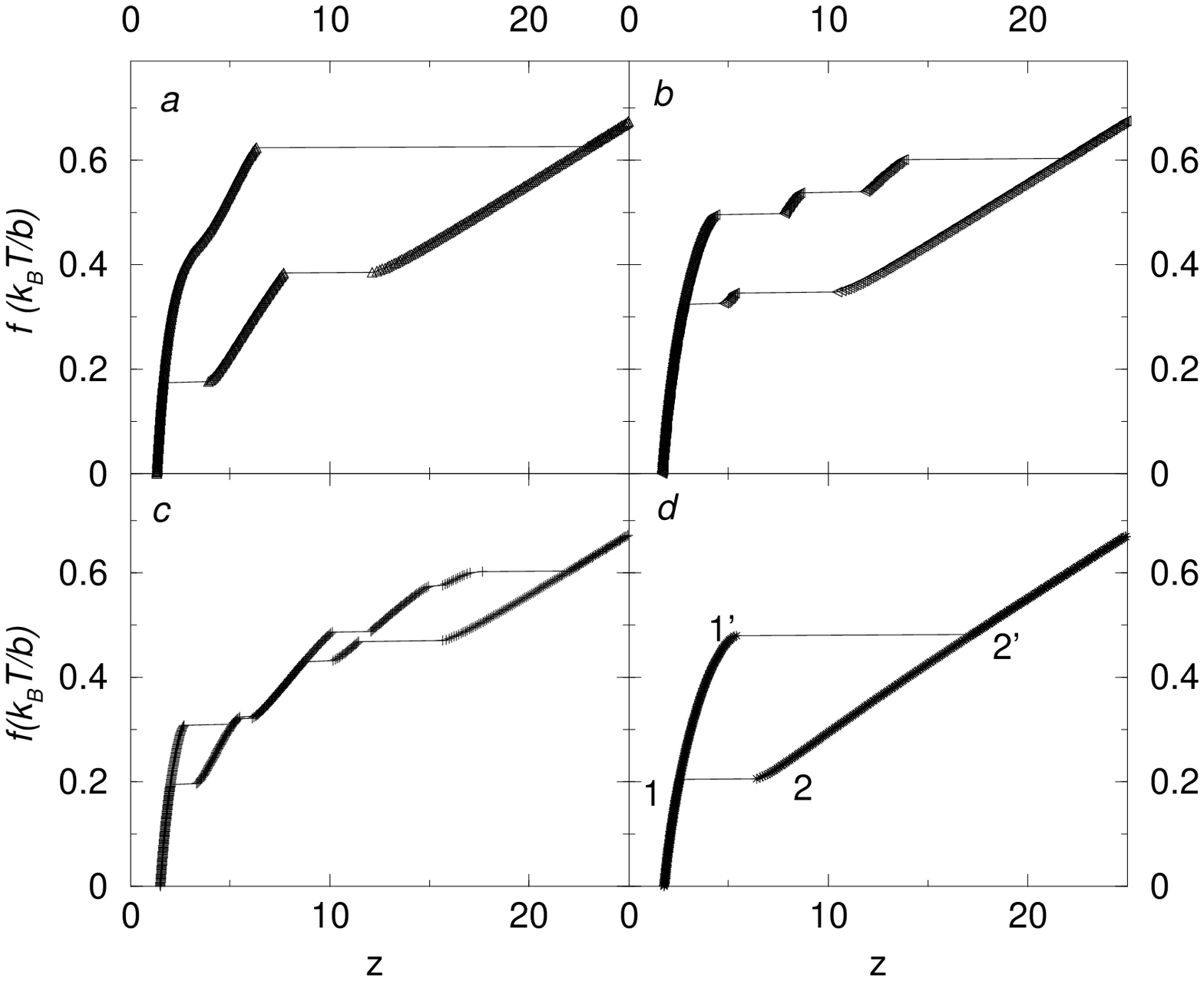}
\caption{Typical force-extension curve 
  which show multiple steps for four different sequences with $\chi/|v_0| =
  0.5$. The hysteresis during stretch-release cycle is shown at each curve.
  The sequence of each polymer is\newline 
a) AAAAB BBBAB BBBBB AABBA BABAA BAABA BAAAB BABAA\newline 
b) BABAA BABBB BBAAA ABBAB BAAAB ABBAA ABBAB BBAAA\newline 
c) ABAAB BAAAA AAAAB BAABB BBAAB BBBAB ABBBB BABAA\newline 
d) AABAB AABBB AABAB BABBB AAABB BABAB ABBAA BABBA }
\label{fig:rco-k05}
\end{figure}

\section{Formation of domains: blocks}

Due to the heterogeneity of the sequences, in average, the structural
transitions are smoothened.  However, unfolding of individual chains shows
unique elastic response reflecting the information in the interactions along
the sequence.  Within a collapsed globular phase, a local sequence of ``n''
monomers has an excess charge with typical fluctuations of size $\sqrt{n}$.
This creates more hydrophobic (type A dominant) cores.  Their sizes
are  determined by the disorder (in local sequence)
and their positions along the
chain are random.  In fig.~\ref{fig:rco-k05}, we show typical examples for the
force-extension relationships which show multiple steps for four different
types of sequences with $\chi/|v_0| = 0.5$.

Multiple plateaus indicate that the  force induced unfolding from 
the globular to
open-string structure occur via intermediate structural forms.  Each plateau
in force-extension relation corresponds to the unfolding of a part of the
chain cooperatively which defines a ``domain''.  The structural changes are
achieved by abrupt unraveling of the corresponding domain. It is clearly
visible that the curves depend on the sequence. For example, curve d)
in  fig.~\ref{fig:rco-k05} corresponds to
the most ``alternating'' sequence (alternating copolymer). Therefore, the
overall poor solvent dominates, and the conformation is a single
globule.  In the
situations  of  curves  a) and d), several A and B blocks are separated by ``non-blocky'' units.
These allow for a more structured phase space, the force-extension relation
is richer. Moreover sequence c) has a net charge 2 (zero charge for
a,b and d).

\begin{figure}
\vspace{5mm} \centering
\includegraphics[width=15cm]{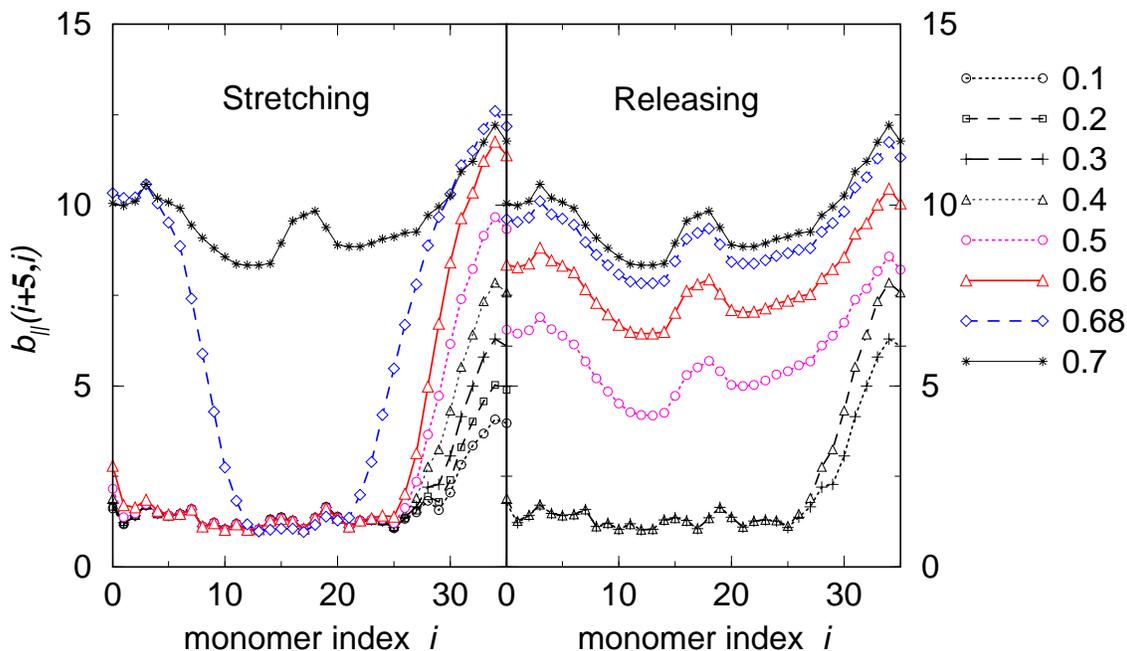}
\caption{The inverse correlation function $b_{\parallel}(i+5,i)$ during the 
  stretch-release cycle.  The chain conformation choose different paths for
  folding and unfolding procedure. The correlation
  $b_{\parallel}(i,j)$ of some part of chain increase/decrease
  cooperatively.  Different symbols  indicate applied force in units 
of $k_BT$.
}
\label{fig:rco-hyscorre}
\end{figure}

In fig.~\ref{fig:rco-hyscorre} the correlation function of a specific sequence
(c in fig.~\ref{fig:rco-k05}) is shown during the stretch-release cycle.  The
regions of small values of $b_{\parallel}(i,j)$ represents the globular phase.
As the force increases, a part of the chain is pulled out from one end leading
to a tadpole like conformation.  Upon further increase of the force, the
remainders in the globular phases are reduced.  Then, the unfolding of the
globule occurs in stepwise patterns, which have also been predicted for
polyelectrolyte necklace chains \cite{vilgis00}. Different configurations are
visited while the chain releases.  The more hydrophobic parts assemble
together at the beginning of the release, the necklace like conformations
are formed.  These clusters are growing and finally merge to a single globule.
The fig.~\ref{fig:rco-hyscorre} also shows that upon releasing the structural
change from necklace to globule occurs at smaller characteristic force than
upon stretching.  The hysteresis on force extension curves is determined by
the energy barriers between two coexisting configurations following a force
induced path. The time for relaxation is exponentially increasing with respect
to the energy barrier $\Delta E^b$ between the two states, $t_r \sim
\exp(\Delta E^b/k_{\rm B}T)$.  These local minima created by quenched disorder
is stable with respect to the fluctuation \cite{shakhnovich89}.  The
hysteresis appears in the experimental situation with finite pulling
speed \cite{kellermayer97}.
 It has  been  already seen that for 
pulling speeds faster  than  $t_r$, the rearrangement of monomers
 at given displacement is not
followed by unfolding of each domain. However, a complete analysis of the
dynamical aspects and the influence of the pulling velocity will be given in a
separate paper \cite{afm01b}.

\begin{figure}
\vspace{5mm} \centering
\includegraphics[width=15cm]{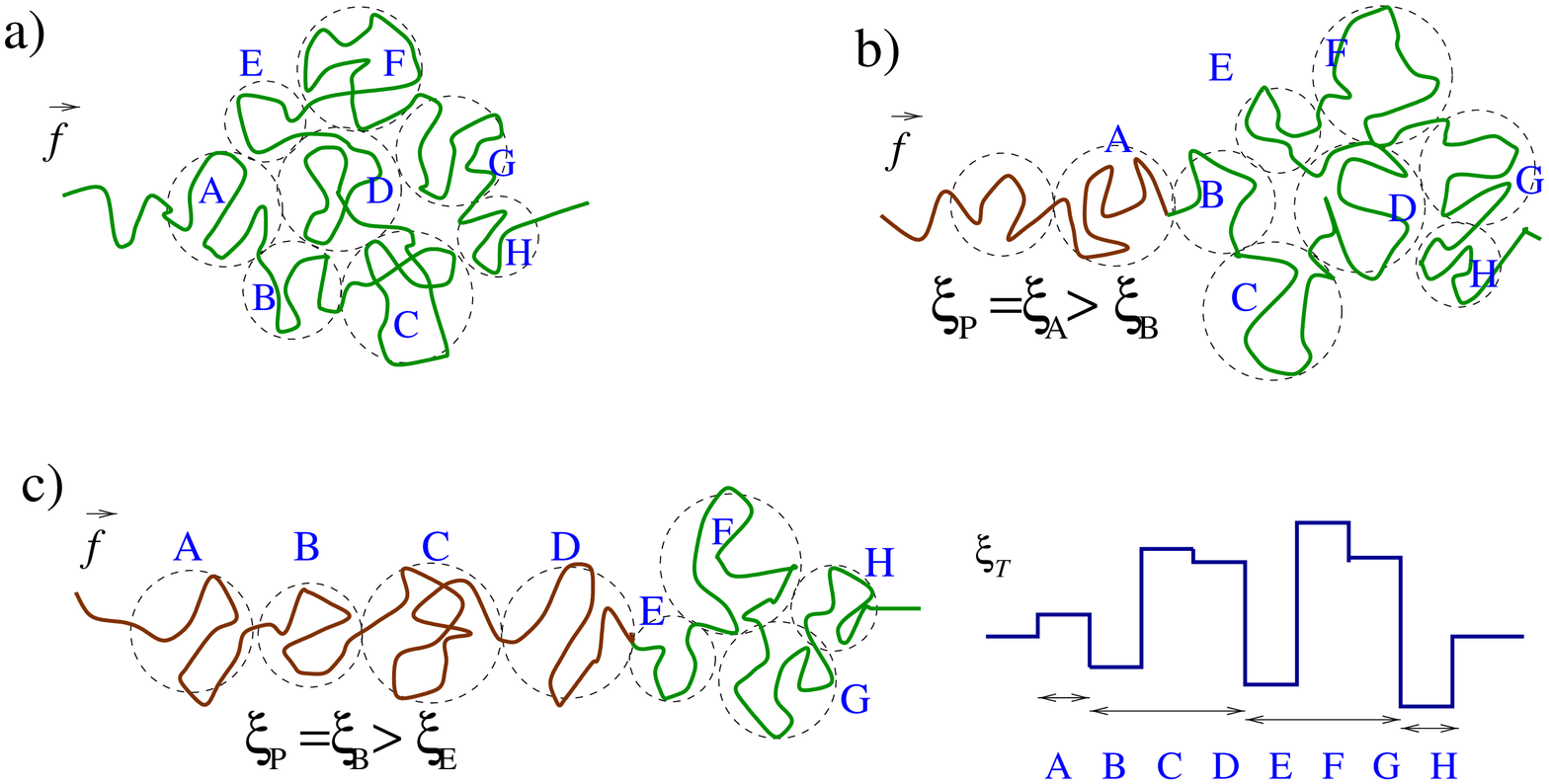}
\caption{Blob interpretation of stretching a random copolymer}
\label{fig:rco-blob}
\end{figure}

The stretching without relaxation is illustrated in fig.~\ref{fig:rco-blob}.
The external force propagates from ends to the center of the chain.  The
extension is controlled by smaller thermal blob size along the path of force
propagation. Variations in thermal blob size $\xi_i$ due to the disorder along
the chain are expected.  Each blob carries the thermal energy $k_{\rm B}T$ and
the variation in size of blob comes from the variation in local sequence.
Applying external force unravels each thermal blob one after another
when the  force reaches the order of $\xi_P = \xi_i$ (i=A,B,C,...). 
The force needed to pull out blob B is large enough to pull blobs C
and D (since $\xi_{\rm C},\xi_D < \xi_B$). 
Therefore, as soon as blob B is released, there is no
further increase of the force required to pull C and D.  This means that the
B-C-D blobs respond to the external force cooperatively.  In similar way, E-F-G
form another domain.  After being pulled out, the  size of each blob 
should be  rescaled to the size of Pincus blob at given force.

In thermodynamic equilibrium at given force $f_0$,  more than one
configuration with different extension can be  accessed  with 
some probability. (For example, 1 and 2 in fig.~\ref{fig:rco-k05} d).  The ratio
between the average life time of two configurations is $P(E_1)/P(E_2)$.
This is related to the free energy difference $\Delta F =-k_{\rm B}T
\ln(P(E_1)/P(E_2))$.  From the force extension curve,  the
free energy difference can be easily obtained: $\Delta F = F_1 - F_2 =
f_0\delta z_{1-2}$.  Estimation of energy barrier along the path 1-1'-2'-2
(shown in  Fig.~\ref{fig:rco-hyscorre} d) is $\Delta E^b_{1-2} =
\int_1^{1'} f(z) dz$.

\section{Conclusion}
We investigated the elastic properties of a single chain with arbitrary sequences
using variational method.  The essential property governing the 
force-extension curve is 
domain-wise unfolding from globular to open string conformation.  The
characteristic force related to each domain is captured as  a ``plateau'' in
the force-extension curve.  In this variational approach, other types of
interactions (including long range interactions) and other properties of the
chain (e.g., stiffness) can be easyly  incorporated.  When long range
interactions are dominant, domains becomes correlated.
This effect will smoothen out the sharp transition due to the unfolding of
each independent domain.  We discuss the role of long range interactions
separately \cite{afm01b}.  Proteins could be also treated in similar way by
assigning special sequence using binary interaction matrix and by adding more
types of monomers which may correspond to the amino acids.  The accuracy of
the method is tested in comparison with Monte Carlo simulation \cite{thijs}.

\acknowledgments
We  thank V. Rostiashvili, S. P. Obukhov and A. Johner 
for thoughtful discussion. N. L. would like to thank T. Vlugt for 
sharing his MC simulation results before the publication.
N. L. Acknowledges the financial
support by the Schwerpunktprogramm Polyelectrolyte of the German Science
Foundation, DFG.


\end{document}